\documentclass[lettersize,journal]{IEEEtran}
\usepackage{amsmath,amsfonts}
\usepackage{algorithmic}
\usepackage{algorithm}
\usepackage{array}
\usepackage[caption=false,font=normalsize,labelfont=sf,textfont=sf]{subfig}
\usepackage{textcomp}
\usepackage{stfloats}
\usepackage{url}
\usepackage{verbatim}
\usepackage{graphicx}
\usepackage{cite}
\usepackage{mathtools}
\usepackage{hyperref}
\usepackage{makecell}
\usepackage{amssymb}
\newtheorem{definition}{Definition}
\newtheorem{proposition}{Proposition}
\newtheorem{lemma}{Lemma}

\begin{document}

    \title{Two Criteria for Performance Analysis of Optimization Algorithms}
    \author{Yunpeng Jing, Hailin Liu,~\IEEEmembership{Senior Staff,~IEEE,} Qunfeng Liu}

    \markboth{Journal of \LaTeX\ Class Files,~Vol.~00, No.~0, August~2024}%
    {Shell \MakeLowercase{\textit{et al.}}: Two Criteria for Performance Analysis of Optimization Algorithms}

    \maketitle

    \begin{abstract}

        Performance analysis is crucial in optimization research, especially when addressing black-box problems through nature-inspired algorithms.
        Current practices often rely heavily on statistical methods, which can lead to various logical paradoxes.
        To address this challenge, this paper introduces two criteria to ensure that performance analysis is unaffected by irrelevant factors.
        The first is the isomorphism criterion, which asserts that performance evaluation should remain unaffected by the modeling approach.
        The second is the IIA criterion, stating that comparisons between two algorithms should not be influenced by irrelevant third-party algorithms.
        Additionally, we conduct a comprehensive examination of the underlying causes of these paradoxes, identify conditions for checking the criteria, and propose ideas to tackle these issues.
        The criteria presented offer a framework for researchers to critically assess the performance metrics or ranking methods, ultimately aiming to enhance the rigor of evaluation metrics and ranking methods.

    \end{abstract}


    \section{Introduction}\label{sec:introduction}
    In the field of optimization, performance analysis—encompassing both the evaluation of individual algorithms and the ranking of multiple algorithms—is fundamental to research.
    This is particularly true for nature-inspired algorithms, where numerous new algorithms have emerged based on various inspirations~\cite{ma2023performance,tang_Review_2021,dokeroglu_Survey_2019 }.
    However, the research on methods for analyzing the performance of these algorithms remains relatively underdeveloped.

    The No Free Lunch Theorem (NFL)~\cite{wolpert_No_1997} presents a significant challenge in the performance analysis of optimization algorithms.
    While nature-inspired algorithms may demonstrate effectiveness on certain benchmarks, the NFL theorem mathematically establishes that all algorithms are equivalent when not considering specific features of the test set.
    This implies that a nature-inspired algorithm can only excel on test functions that share certain characteristics.
    Consequently, two critical questions arise for algorithm analysis: What characteristics of the test problems align with the algorithm's strengths, and why is the algorithm effective for test problems exhibiting those characteristics?
    A crucial prerequisite for addressing these questions is that the evaluation or ranking for the algorithms must be accurate—free from ambiguities and paradoxes.
    This necessity serves as a key motivation and concern of this paper.

    The literature on the performance analysis of optimization algorithms can be categorized into three groups, ordered from application to theory:

    \begin{enumerate}
        \item Papers that propose or utilize performance analysis methods to introduce new algorithms or conduct algorithmic competitions.
        \item Papers that specifically develop new analysis methods, along with related review papers.
        \item Papers examining the theoretical properties of performance analysis, both for the field of optimization and for other fields.
    \end{enumerate}

    In practice, performance analysis of nature-inspired algorithms typically consists of two steps: first, measuring the performance metrics for each algorithm across various test problems, and second, comparing or ranking these algorithms.
    In the first step, a significant issue with performance metrics is that those commonly employed are often related to function values.
    This includes directly using objective function values (best, worst, mean, median, variance)~\cite{elaziz_Hyperheuristic_2019, azizi_Atomic_2021,braik_Chameleon_2021,askari_Heapbased_2020,zhang_Intelligent_,mittal_Meanvarianceskewness_2021}, and counting the number of function evaluations required for the algorithm to achieve a specified level of precision~\cite{abubakar_Hybrid_2022,cantu_Constrainthandling_2021}.
    In response to this scenario, this paper critically analyzes the issues of \emph{function-value based performance metrics}.
    In the second step, researchers typically perform further statistical analyzes using techniques such as Wilcoxon signed ranks test, Kruskal Wallis test, Friedman’s test, Holm’s test, etc.
    These methods are based on average rankings or $p$-values, both of which have notable drawbacks.
    The criticisms against $p$-values are well-documented and widely accepted~\cite{greenland_Statistical_2016,benavoli_Time_,wasserstein_ASA_2016}, while the issues of comparison methods based on average ranking has not been widely discussed.
    In this scenario, this paper focuses on the criticisms of \emph{average ranking based comparison methods}.

    At the methodological level, Review paper~\cite{halim_Performance_2021} comprehensively collects a wide range of performance metrics and statistical methods.
    Many of these performance metrics are based on function values, while many of these statistical methods rely on averaging ranking—both of which are critiqued in this paper.
    In subsection~\ref{subsec:the-examples-of-iia-paradox}, two methods are emphasized as representative examples.
    The first is the non-parametric hypothesis testing method~\cite{garcia_Study_2009}, which, as mentioned earlier, is widely used in current practice.
    The second is Bayesian analysis~\cite{rojas-delgado_Bayesian_2022}, which criticizes the use of $p$-values in the post-hoc procedure of non-parametric hypothesis testing and advocates for Bayesian inference in algorithm comparisons.
    Although both methods use statistical techniques to analyze the experimental results of optimization algorithms, this paper argues that obtaining additional information about the optimization context results in more reliable conclusions than relying solely on experimental data.

    The theoretical layer provides foundational support for algorithmic metrics and ranking methods, such as mathematical proofs of statistical formulas and their properties.
    As previously mentioned, the No Free Lunch (NFL) theorem offers a foundational modeling framework for optimal performance analysis and explains its core principles.
    Another significant direction of research is Liu's exploration of the Condorcet paradox in the comparison of optimization algorithms, leveraging the shared characteristics between algorithm evaluation and voting problems.

    From a macro perspective, this paper identifies three main shortcomings in the existing literature:
    \begin{enumerate}
        \item {The straightforward application of general statistical methods~\cite{garcia_Study_2009,benavoli_Time_,rojas-delgado_Bayesian_2022} overlooks the unique characteristics of evolutionary algorithms in this specific context, leading to less accurate analysis results. Specifically, using statistical methods on experimental data without considering contextual information—such as the characteristics of the objective function—fails to avoid certain logical paradoxes (discussed in detail in section~\ref{sec:the-iia-criterion}).}
        \item {Excessively strong mathematical assumptions hinder the advancement of theoretical conclusions in a way that effectively guides practical application. For instance, the No Free Lunch theorem exemplifies this challenge.}
        \item While merely describing paradoxical phenomena can help identify issues within the field, the absence of deeper theoretical discussions limits guidance on how to address these problems.
    \end{enumerate}

    In response to these challenges, this paper focuses on the unique characteristics of optimization algorithm performance analysis, rather than relying on general statistical or voting theory contexts.
    It capitalizes on the distinct aspects of optimization scenarios in interdisciplinary analyses.
    The paper introduces key concepts while deeply exploring underlying causes, striking a balance between intuitive understanding and practical applicability, all while maintaining a high level of rigor.
    Additionally, it offers practical recommendations and points to future research directions

    This paper specifically investigates two key criteria for the analysis of optimization algorithms.
    The first, the \emph{isomorphism criterion}, stipulates that the evaluation of performance for the same real-world effect should not be influenced by the choice of objective function modeling.
    Performance metrics based on objective function values often violate this criterion.
    The second, the \emph{IIA criterion} (Independence of Irrelevant Alternatives criterion), requires that the comparison results between two algorithms should not be affected by unrelated third-party algorithms.
    Ranking methods that rely on average ranking may potentially violate this criterion.
    Together, these two criteria reflect a common underlying principle: \emph{that algorithm performance should not be influenced by irrelevant factors}.

    The motivation for proposing the principle of isomorphism arises from the observation that the definition of optimization problems in computational optimization often relies on one or more specific functions.
    However, a specific function does not equate to a realistic optimization problem; rather, it serves as a mathematical abstraction of the latter.
    The same realistic optimization problem can correspond to multiple specific functions due to variations in mathematical modeling.
    In essence, a specific function embodies two sets of features: the inherent characteristics of the real-world optimization problem and the model features introduced by the mathematical framework itself.
    Distinguishing model features from the inherent features of a realistic optimization problem is the primary motivation behind the isomorphism criterion.

    The IIA criterion has two sources.
    The first is derived from research in the voting field regarding ranking problems, where the IIA condition from Arrow's Impossibility Theorem~\cite{arrow1951social} is applied to the evaluation of optimization algorithms.
    The second source can be viewed as a symmetrical counterpart to the well-known Simpson's Paradox~\cite{blyth_Simpsons_1972} in statistics.
    When data is organized in a table, with each row representing an object to be compared and each column representing a tester, Simpson's Paradox examines how conclusions change when columns are added or removed, whereas the IIA criterion investigates how conclusions are affected when rows added or removed.

    Detailed discussions of these two criteria will be presented in the following sections: section~\ref{sec:the-isomorphism-criterion} will focus on the isomorphism criterion, and section~\ref{sec:the-iia-criterion} will cover the IIA criterion.
    Each section will include a comprehensive description of the motivation behind the criterion, its definition in both natural and mathematical language, an analysis of the occurrence of the corresponding paradox in practice, an examination of the causes of the paradox, and an exploration of potential solution ideas.
    The final section will summarize the conclusions of the paper, provide specific recommendations, and outline directions for future research.

    \section{The isomorphism criterion}\label{sec:the-isomorphism-criterion}
    This section focuses on the optimization isomorphism criterion.
    First, we provide a detailed explanation of the motivation for studying isomorphism.
    Building on this foundation, we mathematically model the isomorphism of optimization problems and present a formal description of the isomorphism criterion, along with three invariance properties.
    Next, we provide examples related to isomorphism and introduce a theorem for identifying violations of this criterion.
    Finally, we propose a solution concept called isomorphism normalization, which ensures that any performance metric can be adapted to comply with the isomorphism criterion through the application of this normalization process.

    \subsection{The motivation of isomorphism criterion}\label{subsec:the-motivation-of-isomorphism-criterion}
    Typically the most basic form of the optimization problem is expressed as
    \begin{equation}
        \label{eq:optimization}
        \mathop{\arg\min}\limits_{x\in X} f(x)
    \end{equation}
    A wide variety of optimization problems can be derived by adding conditional restrictions or by considering multiple single-objective optimizations together.
    From a purely mathematical standpoint, it may appear that one function corresponds to a specific optimization problem, while different functions represent different optimization problems.
    However, a critical aspect often overlooked by researchers is that, in the context of real-world issues, an objective function is merely one of many possible abstract approaches to addressing a specific optimization problem.
    The essence of the optimization problem itself does not rely on this particular function.
    This oversight leads to a subtle issue: scholars studying optimization problems based on a specific objective function may become unduly influenced by the unique characteristics of that function.

    \begin{figure}[!t]
        \centering
        \includegraphics[width=3.5in]{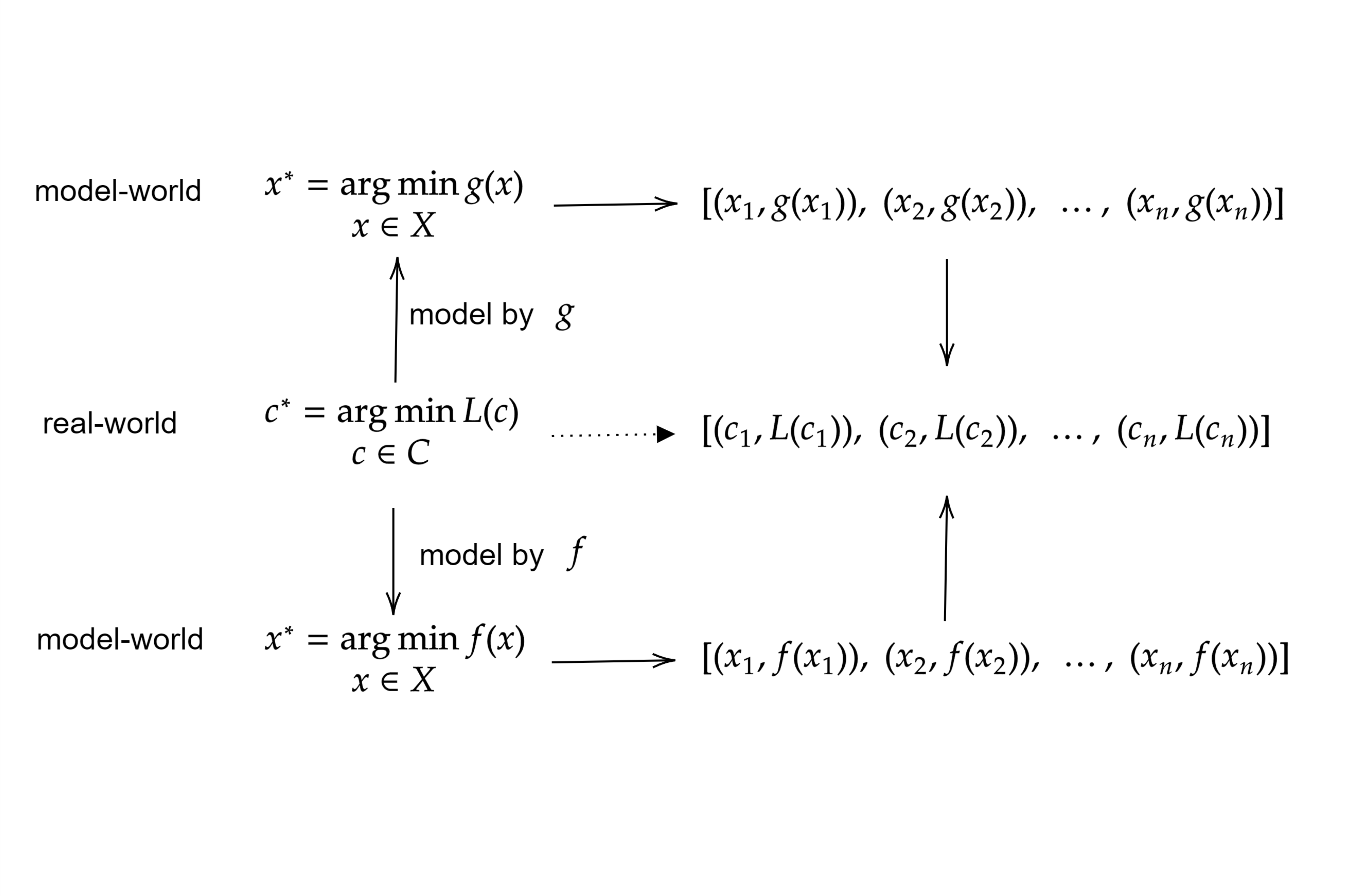}
        \caption{Real-world problem and functional problems.
        A real-world search process produces the same actual results; however, when the problem is modeled using different functions, the function values may differ. Despite this, because the real-world results are the same, the performance evaluations should remain invariant.}
        \label{fig:fig_real_world}
    \end{figure}

    As illustrated in Fig.~\ref{fig:fig_real_world}, consider a real-world problem that requires identifying the best decision $c^*$ from a decision set $C$, with the real-world cost of decision $c$ denoted as $L(c)$, and the problem is denoted as $(X, L)$,
    \begin{equation}
        \label{eq:realworldop}
        \mathop{\arg\min}\limits_{c\in C}L(c)
    \end{equation}

    It is important to note that $c$ and $L$ are not mere numbers; they represent real-world decisions and costs.
    Suppose two researchers aim to solve the problem $(C,L)$.
    The first researcher abstracts this problem into a mathematical problem $(X,f)$, where each $c \in C$ corresponds exactly to a mathematical object $x \in X$, and each $L(c)$ corresponds to a specific $f(x)$, which is an objective function value.
    After modeling, the researcher uses an algorithm to solve the problem $(X,f)$, yielding a search sequence $s=(x_1, x_2, \dots, x_n)$ and the corresponding function values $(f(x_1),f(x_2),\dots, f(x_n))$.\@
    The performance evaluation of this process is denoted as $P_{s,f}$.
    The second researcher follows a similar approach but uses a different modeling function $g$, resulting in the same search sequence $s$.\@
    Therefore, the performance evaluation of this process can be represented as $P_{s,g}$.
    Now we pose the question: should $P_{s,f}$ equal $P_{s,g}$?

    From a real-world perspective, the answer is clear.
    Both processes address the same real-world problem, generate the same decision suggestions, and result in identical real-world outcomes.
    Thus, their performance evaluations—regardless of the metrics used—should be the same.

    We refer to the relationship between different mathematical models that can represent the same real-world optimization problem as optimal isomorphism, or simply isomorphism when there is no ambiguity.
    Consequently, the aforementioned criterion can be termed the isomorphism criterion, which can be articulated in natural language as follows:

    \begin{definition}[isomorphism criterion]
        The evaluation of a real-world process's performance should remain consistent, regardless of the mathematical model used.
    \end{definition}

    In the following two subsections, we provide a precise mathematical formulation to enable a more in-depth analysis of isomorphism.

    \subsection{The model of isomorphism}\label{subsec:the-model-of-isomorphism}

    From the form of Equation~\ref{eq:optimization}, the equivalence of two functions in terms of “optimization” can be defined as follows,

    \begin{definition}
    [isomorphism]
        \label{def:iso-def}
        A function $f$ is said to be isomorphic to a function $g$ if and only if $\forall X$,
        \begin{equation}
            \label{eq:iso-def}
            \mathop{\arg\min}\limits_{x\in X} f(x) = \mathop{\arg\min}\limits_{x\in X} g(x),
        \end{equation}
        and it denoted as $f\sim g$.
    \end{definition}

    Since functions that are isomorphic to each other must share the same feasible domain, this paper omits the specification of the domain and assumes that all discussions are based on the premise that the independent variables lie within identical feasible domains.
    In addition, although the return value of an optimization problem should theoretically be considered a solution set, this paper treats the optimization problem as if it returns a single solution for three reasons.
    First, this approach minimizes lengthy categorical discussions.
    Second, all descriptions, proofs, and related elements can be easily adapted by replacing the single solution with a set.
    Finally, the primary purpose of this paper is to provide practical guidance rather than engage in strict mathematical discourse.
    Consequently, the subsequent sections will reference these descriptions to minimize unnecessary redundancy.

    Definition~\ref{def:iso-def} emphasizes that the optimal solution
    remains unchanged regardless of the model used, a concept referred to as the \emph{optimal invariance}.

    Using this definition, we introduce two necessary and sufficient conditions for studying the nature and determination of isomorphism.
    The discussions derived from these two propositions demonstrate that isomorphism is a common phenomenon in optimization fields.

    \begin{proposition}[comparison invariance]
        \label{prop:ci}
        $f\sim g$ if and only if $\forall x_1, x_2$, $sign(f(x_1)-f(x_2))=sign(g(x_1)-g(x_2))$, where $sign$ is the signum function.
    \end{proposition}
    \begin{IEEEproof}
        Necessity: Let $X=\{x_1, x_2\}$, it is clear that the equation holds by the definition of isomorphism.
        Sufficiency: Using the method of contradiction, assume there exists a set $X$ such that the minimum values of $f$ and $g$ do not coincide, and denote these two minimum values as $x_1$ and $x_2$, respectively.
        A contradiction can then be easily derived, making sufficiency evident.
        In summary, the proposition holds.
    \end{IEEEproof}

    Clearly, comparison invariance can be extended to mean that when examining any finite set of points on two mutually isomorphic functions, the ranking of the corresponding function values should remain identical.
    This implies that an algorithm will generate the same search sequence when dealing with isomorphic optimization functions, provided its search strategy relies solely on comparisons.
    For instance, if $f \sim g$ and both functions are evaluated separately using the PSO algorithm, the individual positions, velocities, historical optimas, and global optima will maintain the same in each generation, as long as the initial population is chosen consistently and a fixed sequence of random numbers is used.

    \begin{proposition}[monotonic invariance]
        $f\sim g$ if and only if there exists a monotonically increasing function $h$, such that $f=h\circ g$.
    \end{proposition}

    This proposition is clearly established by considering Proposition~\ref{prop:ci} as well as the definition of monotonically increasing functions.

    Compounding the original optimization objective function with a monotonically increasing function is a common practice in solving optimization problems.
    For instance, taking the logarithm of an expression is standard when estimating parameters in a distribution using the maximum likelihood method.
    Another more common form of composite is the linear composite, where $h$ is a primary function with a positive coefficient.
    Consequently, we can define the linear isomorphism as
    \begin{definition}[linear isomorphism]
        A function $f$ is said to be linearly isomorphic to a function $g$ if and only if there exists a positive real number $a$ and a real number $b$ such that $f=a*g+b$.
    \end{definition}
    This relation is denoted as $f \overset{L}{\sim} g$.

    It is common to model the same real-world problem with different functional expressions that are linearly isomorphic; for instance, calculating the minimum temperature using Fahrenheit, Celsius, and Kelvin results in distinct functions, all of which are linearly isomorphic.
    More examples can be seen in subsection~\ref{subsec:isomorphism-analysis-for-metrics-in-practice}.

    \subsection{The isomorphism criterion}\label{subsec:the-isomorphism-criterion}

    The definition of isomorphism lays the groundwork for a mathematical description of the isomorphism criterion.
    To provide a more comprehensive understanding of this criterion, we first define the isomorphism set and the performance metric.

    The relation $\sim$ is indeed an equivalence relation, as it can be shown to be self-reflexive, symmetric, and transitive based on its definition.
    The equivalence set derived from this relation is referred to as the isomorphism set, and denoted as $I(f)$.

    \begin{definition}[isomorphism set]
        \begin{equation}
            \label{eq:isoset}
            I(f) \coloneqq \{g| g\sim f\}
        \end{equation}
    \end{definition}

    Without loss of generality, we denote the search sequence as $s = (x_1, x_2, \dots, x_n)$ and the metric as $M$.\@
    If the calculation of $M$ involves corresponding function values, $M$ can be viewed as a generalized function parameterized by the objective function $f$, and the result $M(f)$ is a function that parameterized by $s$.\@
    Thus, the complete computational procedure can be represented as $M(f)(s)$.

    Then, the mathematical form of isomorphism criterion can be represented as,

    \begin{definition}[isomorphism criterion]
        A metric $M$ is said to be consistent with the isomorphism criterion if and only if $\forall g\in I(f), M(g)\equiv M(f)$.
    \end{definition}

    The symbol $\equiv$ implies that $\forall s=(x_1, x_2, \dots, x_n), M(f)(s)=M(g)(s)$, and a metric that is consistent with the isomorphism criterion is called a isomorphic metric.
    Similar concepts can be similarly applied to linear isomorphisms and are not necessarily redundantly defined in this context.

    \subsection{Isomorphism analysis for metrics in practice}\label{subsec:isomorphism-analysis-for-metrics-in-practice}
    Traditional optimization algorithms, like Newton's method, focus on finding locally optimal solutions by utilizing the gradient information of a function.
    This approach enables the search sequence to move incrementally closer to the optimal solution, thus giving meaning to the order of convergence $q>1$ and the rate of convergence $\mu$ in the following equation:
    \begin{equation}
        \label{eq:convrate}
        \lim_{n\to\infty}\frac{|x_{n+1}-L|}{|x_{n}-L|^q}=\mu
    \end{equation}
    where $x^*$ is the local optimal solution and $\mu$ is a constant.

    Equation~\ref{eq:convrate}  is clearly related solely to the search sequence, thereby adhering to the isomorphism criterion.
    Since it focuses on the convergence behavior independent of the specific objective function, it maintains invariance across isomorphic functions.

    However, in nature-inspired algorithms aimed at finding the global optimal solution for black-box problems, the search sequence does not ensure that the current solution progressively approaches the global optimum.
    This makes it challenging to directly apply Equation~\ref{eq:convrate} for measuring the algorithm's convergence speed, as the behavior of the search can be more erratic and less predictable.

    A conceded scheme~\cite{dhivyaprabha_Synergistic_2018} for evaluating the convergence rate of a nature-inspired algorithm is shown in Equation~\ref{eq:convrate2}, and similar variants~\cite{liu_Convergence_2017,he_Average_2016, sadollah_Mine_2018} will not be enumerated here because they share the same issues.
    \begin{equation}
        \label{eq:convrate2}
        conv.rate=\frac{|f_i-f^*|}{|f_{i-1}-f^*|},
    \end{equation}
    where $f^*$ is the global optimum of the function and $f^i$ is the optimum that can be reached after $i$ iterations.

    When substituting $f$ into Equation~\ref{eq:convrate2} with $g=af+b$, we observe that the value of Equation~\ref{eq:convrate2} remains unchanged because subtraction eliminates $b$ and division cancels out $a$.\@
    Consequently, the metric described by Equation~\ref{eq:convrate2} is linearly isomorphic.
    However, when examining other isomorphic functions, such as $g=f^3$ or $g=\ln(f-f^*+1)$, Equation~\ref{eq:convrate2} cannot guarantee the same evaluation for the same search sequence, which means that the metric is not isomorphic.

    Another example of non-isomorphic metric is the Inverted Generational Distance (IGD)~\cite{qingfuzhang_RMMEDA_2008}, which is a commonly used metric in multi-objective optimization, and is defined as follows,
    \begin{equation}
        \label{eq:igd}
        D(P^*,P)=\frac{\sum_{v\in P^*}d(v,P)}{|P^*|}
    \end{equation}
    where $P^*$ is a set of uniformly distributed points in Pareto Front(PF) or Pareto Set(PS), $P$
    is an approximation to the PF or PS, and $d(v,P)$ is the minimum distance between $v$ and the points in $P$.

    There are two versions of IGD: IGDX~\cite{aiminzhou_Approximating_2009} in the decision space, which is consistent with the isomorphism criterion, and IGDF~\cite{aiminzhou_Approximating_2009} in the objective space, which is clearly not consistent with the isomorphism criterion.

    The above examples show that there are metrics being used that violate the isomorphism criterion, but at the same time, there are metrics that meet the isomorphism criterion, or, more weakly, the linear isomorphism criterion.

    Although Equation~\ref{eq:convrate2} does not explicitly define the phenomenon of linear isomorphism, we consider the linear isomorphism criterion to be a widely accepted consensus.
    To the best of our knowledge, the earliest papers to explicitly address this phenomenon (termed strong homogeneity in their work) were~\cite{elsakov_Homogeneous_2010} and~\cite{zilinskas_Strong_2012}.
    This raises two questions:
    \begin{enumerate}
        \item Why has the isomorphism criterion not been widely discussed?
        \item Why should the isomorphism criterion be emphasized now?
    \end{enumerate}
    These two questions address why isomorphism matters, and the discussions will be elaborated in the next subsection.

    \subsection{Why does isomorphism matter?}\label{subsec:why-does-isomorphism-matter?}
    There are three possible reasons why the isomorphism criterion is not widely discussed.
    First, many scholars focus on function problems as the subject of optimization research, overlooking the modeling process that transforms real-world problems into function problems.
    Second, when researchers aim to propose new algorithms, the analysis often narrows to demonstrating that their algorithms outperform others, making the accuracy of performance evaluation less critical as long as comparisons can be made.
    Third, while the existence of the isomorphism problem is acknowledged and solutions for linear isomorphism are proposed, finding a universal solution for all isomorphism problems remains challenging, leading to the criterion being sidelined.

    The analysis of the above three possible reasons constitutes the answer to the question of why the isomorphism criterion needs to be emphasized now.

    First, this paper illustrates through Figure~\ref{fig:fig_real_world} that the same real-world problem can lead to multiple modeling approaches that are equivalent in optimization but represented by different specific functions.
    Modeling is an artificial construct designed to address real-world issues, which reflects the essence of the optimization problem while also introducing additional functional characteristics.
    The primary goal of introducing the concept of isomorphism is to delve into these fundamental aspects and eliminate the extraneous features.

    Second, this paper advocates for a shift in optimization research from merely proposing new algorithms to analyzing the underlying mathematical principles of nature-inspired algorithms.
    In this new focus, accurate performance evaluation becomes crucial, making the isomorphism criterion highly significant.

    Finally, by explicitly defining and analyzing isomorphism, it becomes possible to construct metrics that are applicable to all isomorphisms, as explored in the following two subsections.

    \subsection{A necessary condition for isomorphism metrics}\label{subsec:a-necessary-condition-for-isomorphism-metrics}
    Intuitively, whenever the definition of a metric $M$ uses the value of a function, then the metric is rightfully affected by the model.
    However, this characterization is not rigorous.
    For example, $M = sign(f(x_1)-f(x_2))$ is an expression that contains a part that calls the value of the objective function and that also satisfies the isomorphism criteria.
    This example leads us to think about whether there is a more specific determination condition that can be used to judge that a metric violates the isomorphism criterion

    To describe this condition, we first treat the expression $M(f)(s)$ as $T(u_1, u_2, \dots, u_n)$ ,where $u_i=f(x_i)|_1^n$, and introduce the following lemma.

    \begin{lemma}
        \label{lemma:l1}
        Provided that there is an objective function $f$ and a search sequence $s=(x_1, x_2, \dots, x_n)$, holds $f(x_1)>f(x_2)>\dots>f(x_n)$, for all $j\in\{1, 2, \dots, n\}$, there exists a real number $t>0$, such that $\forall 0<\varepsilon<t, \exists g\in I(f)$, such that $g(x_j)=f(x_j)+\varepsilon$ and $g(x_i)=f(x_i)$ where $i \in \{1, 2, \dots, n\}\backslash j$.
    \end{lemma}
    \begin{IEEEproof}
        When $j=1$, let $t$ be any positive number, otherwise, let $0<t<f(x_{j-1})-f(x_j)$, so that by the definition of isomorphism the lemma is proved.
    \end{IEEEproof}
    \begin{proposition}
        \label{prop:iso}
        A necessary condition for a metrics $M$ to satisfy the isomorphism criterion is that the corresponding function $T$ holds $\frac{\partial T}{\partial u_i}=0|_1^n$, whenever $\frac{\partial T}{\partial u_i}$ exists.
    \end{proposition}
    \begin{IEEEproof}
        Since $\frac{\partial T}{\partial u_i}$ exists,
        \begin{equation}
            \label{eq:dtdui}
            \frac{\partial T}{\partial u_i} = \lim\limits_{\varepsilon\to 0}\frac{T(u_, \dots, u_i+\varepsilon, \dots, u_n) - T(u_1, \dots, u_n)}{\varepsilon}.
        \end{equation}
        Without losing generality, we suppose that $u_1>u_2>\dots>u_n$.
        Considering Lemma~\ref{lemma:l1}, there exists functions $f$ and $g$ holds $f\sim g$, such that $T(u_1, u_2, \dots, u_j+\varepsilon, \dots, u_n)=M(g)(s)$ and $T(u_1, u_2, \dots, u_n)=M(f)(s)$.
        Considering the definition of isomorphism criterion, we have $\frac{\partial T}{\partial u_i}=0$, then the proposition is proved.
    \end{IEEEproof}

    It is straightforward to determine that Equations~\ref{eq:convrate2} and~\ref{eq:igd} do not satisfy the isomorphism criterion using Proposition~\ref{prop:iso}.

    When only one particular isomorphism is considered, such as linear isomorphism, a metric similar to Equation~\ref{eq:convrate2} can be constructed based on the corresponding characteristics.
    However, when all possible isomorphisms are considered, in conjunction with Proposition~\ref{prop:iso}, constructing a reasonable and simple $T$ is very difficult.
    On the other hand, as analyzed before, the function expression itself carries two parts of information: one part is the information that can reflect the nature of the optimization problem, which is shared in its corresponding isomorphic set, and it will produce the same comparison result when comparing different decisions.
    The other part is the information that is specific to this function itself, which produces a different function value when the function is called.
    As long as we can find the part of information that represents the essence, we can solve the isomorphism paradox.
    More discussions are expanded in the next subsection.

    \subsection{Isomorphism normalization: a solution idea}\label{subsec:isomorphism-normalization:-a-solution-idea}

    Proposition~\ref{prop:iso} demonstrates the difficulty of using conventional means to construct an indicator $M$ that satisfies the isomorphism criterion.
    When we restrict the data source for analyzing performance to experimental data results, these values are inherently attached to the characteristics of the objective function, and while some technical means can partially eliminate the isomorphism problem, Proposition~\ref{prop:iso} demonstrates completely eliminating isomorphism or making the construction of $T$ is very complicated.
    It is difficult and unreasonable to eliminate modeling effects using finite data that have already been subjected to modeling effects.

    A very important point made in this paper is that the analysis of algorithm performance should not restrict its focus exclusively to experimental numerical results.
    Although nature-inspired optimization algorithms are only allowed to obtain information about the objective function information by calling the objective function during the computation process(which is called black-box), it is possible (and highly recommended) to make use of the white-box information of the objective function during the performance analysis.
    Based on this idea, we propose the notion of isomorphic normalization,

    \begin{definition}
        A function $N:\mathbb{F}\to\mathbb{F}$ is an isomorphism normalization function if and only if it has the following properties:
        \begin{enumerate}
            \item $N(f)$  exists for any bounded function $f$.[\emph{existence}]
            \item $N(f)\sim f$.[\emph{isomorphism}]
            \item $N(f)=N(g)\Leftrightarrow f\sim g$.[\emph{uniqueness}]
        \end{enumerate}
    \end{definition}
    where $\mathbb{F}$ is the set of all bounded functions.

    For any performance metric $M(f)$ which does not conform to the criterion of isomorphism, its objective function can be replaced by the one after isomorphism normalized by $N$ to obtain the metric $M'(f)=M(N(f))$, which is obviously conform to the isomorphism criterion because of the uniqueness property of $N$.
    Since isomorphic normalization maps all functions in the same isomorphism set to a single function, this operation is considered to eliminate the differences between the different functions in the isomorphism set and preserve the essence of the corresponding optimization problem.

    An obvious conclusion is that if a function $N$ is an isomorphism normalization function, then $N' = h(N)$, obtained after compounding $N$ with a monotonically increasing function $h$, can likewise satisfy the above three properties, and thus isomorphism normalizations are not unique.

    \section{The IIA criterion}\label{sec:the-iia-criterion}
    This section focuses on the IIA criterion for optimization.
    Two motivations for the IIA criterion are discussed, the definition of IIA criterion in the context of optimization scenarios is defined, and it is verified that the typical statistical approach suffers from the IIA paradox.
    Then, we analyze the reasons for this paradox and propose corresponding solution ideas based on them.

    \subsection{The motivations of IIA criterion}\label{subsec:the-motivations-of-iia-criterion}

    If we analogize algorithms to candidates, test problems to voters, and performance results to voter preference, then comparisons of algorithms based on performance data can be likened to voting problems.
    This equivalence has two aspects, facing similar paradoxes and governed by the same theoretical framework.
    Liu~\cite{liu_Paradoxes_2020,liu_Meanbased_2024} introduces the Condorcet paradox, which is one of the paradoxes in voting theory, to validate the performance comparisons of algorithms.
    This paper further explores the latter aspect.

    In voting theory, Arrow's Impossibility Theorem~\cite{arrow1951social} presents a challenge similar to the No Free Lunch theorem, asserting that no voting method can effectively aggregate individual preferences into a collective conclusion while satisfying a set of plausible criteria.
    This includes the IIA criterion, which can be applied to algorithmic comparison.
    This paper emphasizes the IIA criterion for two reasons: it is less recognized in performance analysis, and comparing optimization algorithms differs from voting because algorithmic data includes specific quantifiable values.
    This additional information provides an opportunity to convert the “impossible” into the “possible.”

    \begin{figure}[!t]
        \centering
        \includegraphics[width=3.5in]{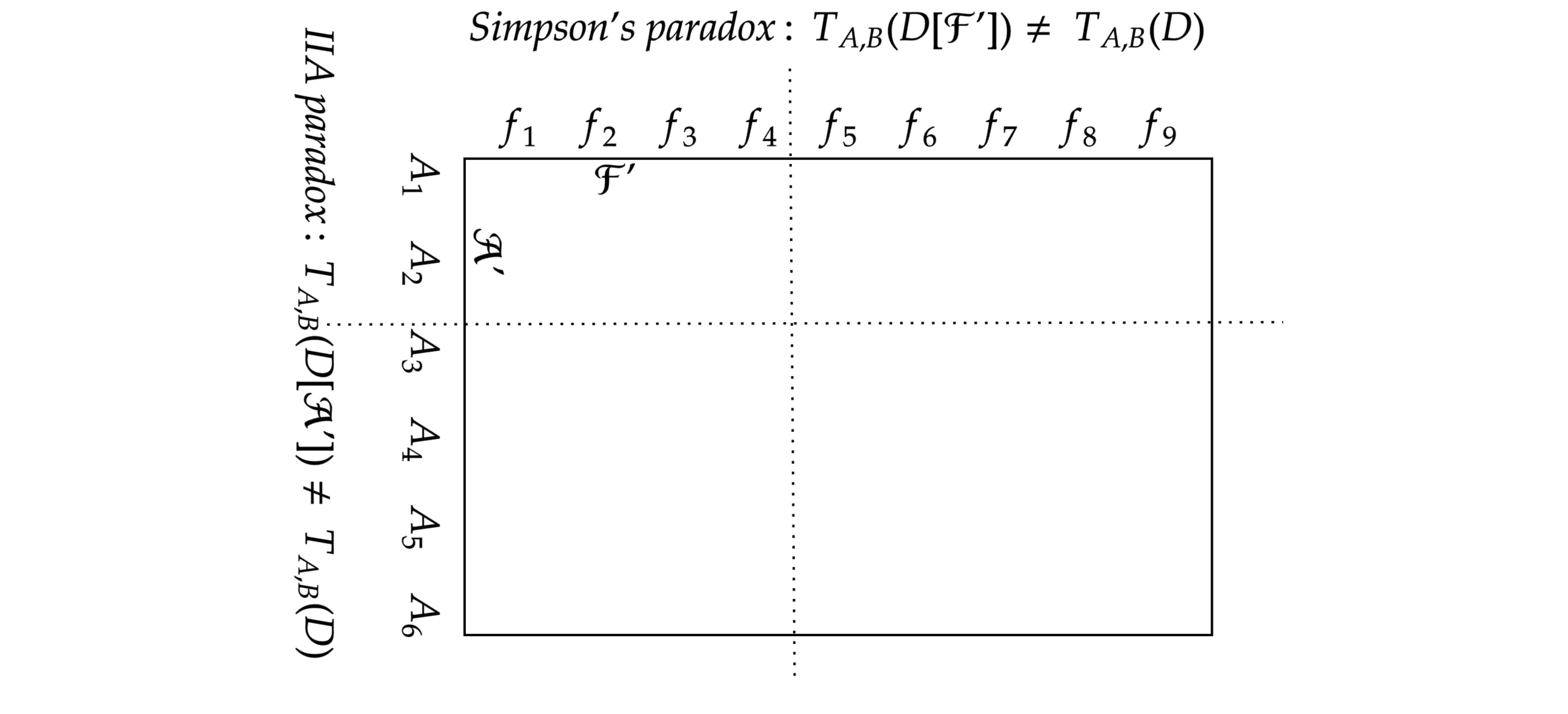}
        \caption{IIA Paradox and Simpson's Paradox. Simpson's paradox emphasizes that comparison results are influenced by the set of benchmark functions, and IIA paradox symmetrically emphasizes that comparison results are influenced by the set of algorithms.}
        \label{fig:iia-simpson}
    \end{figure}

    Without introducing voting theory and Arrow's Impossibility Theorem, the IIA criterion can also be understood from another perspective.
    In statistical theory, there is a well-known Simpson's Paradox~\cite{blyth_Simpsons_1972}, which highlights that different conclusions can arise from using various test groups to analyze the same subjects.
    Simpson's Paradox demonstrates that differing conclusions can emerge under fixed subjects.
    Symmetrically, the IIA paradox reflects that different conclusions can arise from varying subjects while keeping the testers constant.

    \subsection{The IIA criterion}\label{subsec:the-iia-criterion}

    \begin{figure}[!t]
        \centering
        \includegraphics[width=3.5in]{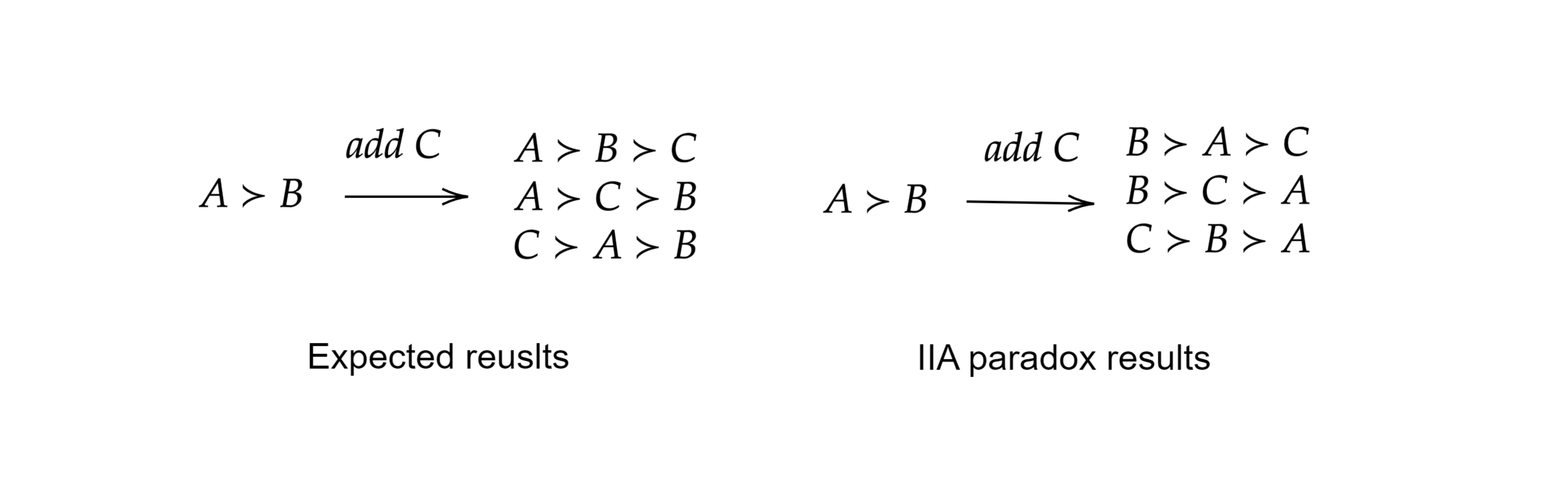}
        \caption{A schematic diagram illustrating the IIA paraodx.
        The irrelevant algorithm $C$ unexpectedly influenced the comparison result of $A$ and $B$.}
        \label{fig:iia}
    \end{figure}
    Fig.~\ref{fig:iia} illustrates the IIA paradox in algorithm benchmarking.
    When we compare algorithms $A$ and $B$, and obtain a result, let's assume $A\succ B$.
    Then, when we introduce algorithm $C$ into the algorithm set, and provided that the analysis method is accurate, we would expect one of the following outcomes: $C\succ A\succ B$, $A\succ C\succ B$, or $A\succ B\succ C$.
    However, if the results show that the addition of $C$ results in $B \succ A$, we can say that the analysis method exhibits the IIA paradox.
    Below, we provide a formal symbolic definition of the IIA paradox .

    A succinct version of the IIA criterion can be described as follows:

    \begin{definition}
        The results of a performance comparison between two algorithms should not be influenced by unrelated third-party algorithms.
    \end{definition}

    More specifically, let $D$ be the original performance data of algorithm set $\mathcal{A}$ on some test set $\mathcal{F}$, $T$ be a data analysis method, and $T_{AB}(D)$ denotes the performance comparison result of algorithm $A$ and algorithm $B$ after analyzing the data $D$, where $\{A,B\}\subset \mathcal{A}$.\@
    Suppose some algorithms other than $A$ and $B$ in $\mathcal{A}$ are deleted from $\mathcal{A}$ to obtain a new set of algorithms $\mathcal{A}' \subsetneq A$, and accordingly the data of these algorithms is deleted from the set $D$ to obtain a projection $D[A']$, then the IIA criterion can be described more precisely as:

    \begin{definition}
        A comparison method $T$ is said to be consistent with the IIA criterion if and only if $\forall \mathcal{A}'\subsetneq \mathcal{A}$, holds
        \begin{equation}
            \label{eq:iia}
            T_{A,B}(D[\mathcal{A}']) = T_{A,B}(D).
        \end{equation}
    \end{definition}

    Figure~\ref{fig:iia-simpson} can be used as an aid to make this formula and the relationship between the IIA paradox and Simpson's paradox easier to understand, in which Simpson's paradox can be described as $ T_{A,B}(D[\mathcal{F}']) \neq T_{A,B}(D)$, where $D[\mathcal{F}']$ denotes the projection of the data $D$ onto $\mathcal{F}' \subsetneq \mathcal{F}$.

    \subsection{The examples of IIA paradox}\label{subsec:the-examples-of-iia-paradox}

    The IIA criterion appears reasonable; however, the statistical methods used in practice do not satisfy this criterion.
    To illustrate this phenomenon concretely, this subsection designs a dataset $D$ and analyzes it using both nonparametric hypothesis testing and Bayesian analysis.
    The results indicate that the IIA paradox is indeed possible.

    Provided that there are 500 problems and 100 algorithms.
    In this context, we designate two specific algorithms as $A$ and $B$, while the remaining algorithms are labeled as $C_1$ through $C_{98}$.
    For the first 100 problems, we assume the following performance results: Algorithm $A$ achieves a score of 1, Algorithm $B$ obtains a score of 99, Algorithm $C_1$ receives a score of 100, and the score of $C_i|_2^{98}$ is $i$.
    For the subsequent 400 problems, we assume the following performance results : Algorithm $A$ is assigned a score of 99, Algorithm $B$ is attributed a score of 98, Algorithm $C_1$ maintains a score of 100, and for any $C_i$ (where $i$ ranges from 2 to 98), its score becomes $i-1$.

    The performance data $D$ can be represented by the matrix as follows,
    \begin{equation}
        \label{eq:data2}
        \bordermatrix{
            & I_1 & I_2 & \cdots & I_{100} & I_{101} & I_{102} & \cdots & I_{500}  \cr
            A & 1 & 1 & \cdots & 1 & 99 & 99 & \cdots & 99 \cr
            B & 99 & 99 & \cdots & 99 & 98 & 98 & \cdots & 98\cr
            C_1 & 100 & 100 & \cdots & 100 & 100 & 100 & \cdots & 100 \cr
            C_2 & 2 & 2 & \cdots & 2 & 1 & 1 & \cdots & 1 \cr
            C_3 & 3 & 3 & \cdots & 3 & 2 & 2 & \cdots & 2 \cr
            \vdots & \vdots & \vdots & \vdots & \vdots & \vdots &\vdots & \vdots & \vdots \cr
            C_{97} & 98 & 98 & \cdots & 98 & 97 & 97 & \cdots & 97 \cr
        }.
    \end{equation}

    Now let $\mathcal{A}'=\{A, B, C_1\}\}$, then $D[\mathcal{A}']$ can be represented as:

    \begin{equation}
        \label{eq:data1}
        \bordermatrix{
            & I_1 & I_2 & \cdots & I_{100} & I_{101} & I_{102} & \cdots & I_{500}  \cr
            A & 1 & 1 & \cdots & 1 & 99 & 99 & \cdots & 99 \cr
            B & 99 & 99 & \cdots & 99 & 98 & 98 & \cdots & 98\cr
            C_1 & 100 & 100 & \cdots & 100 & 100 & 100 & \cdots & 100
        }.
    \end{equation}

    First, we analyze these two datasets separately using non-parametric hypothesis testing method.

    Friedman's test is adopted to assess overall differences, and the Bonferroni-Dunn's procedure is used to calculate the critical difference (CD) values.
    The results are presented in Table~\ref{tab:npht}.
    For data $D$, the $p$-value obtained from Friedman's test falls below the significance level of $\alpha=0.001$, signifying a significant difference in algorithm performance.
    The difference in the average rankings between Algorithm $A$ and Algorithm $B$ exceeds their respective CD value, which establishes the statistical significance of $A\succ B$.
    However, when the same data and analytical process were applied to $D[\mathcal{A}']$, a contrary conclusion was reached, which is also deemed statistically significant.

    \begin{table}
        \begin{center}
            \caption{Result of non-parametric hypothesis testing.}
            \label{tab:npht}
            \begin{tabular}{c|cccccc}
                \hline
                & \makecell{Friedman \\ p-value} & \makecell{critical \\ difference\\($\alpha=0.001$)}               & \makecell{average       \\ rank               \\ of A}               & \makecell{average       \\ rank               \\ of B}               & result       \\
                \hline
                $D[\mathcal{A}']$               & 3.94e-183       & 0.227 & 1.8 & 1.2 & $B\succ A$ \\
                $D$               & 0.0       & 8.105 & 79.4 & 98.2 & $A \succ B$ \\
                \hline
            \end{tabular}
        \end{center}
    \end{table}

    From these results, it can be seen that analysis based on the average rank may lead to the IIA paradox, even when these results are considered statistically significant under non-parametric hypothesis testing.

    Then, we explore the scenario of using Bayesian analysis.
    The experiment adopts the Bradley-Terry model, uses Zermelo's algorithm along with Newman's improved formula to calculate the parameters.

    The results are presented in Table~\ref{tab:bayes}, which reveals that Bayesian inference also encounters the IIA paradox.
    \begin{table}
        \begin{center}
            \caption{Result of Bayesian analysis.}
            \label{tab:bayes}
            \begin{tabular}{c|cccccc}
                \hline
                & $\theta_A$ & $\theta_B$ & $P(A\succ B)$ & $P(B\succ A)$ & result     \\
                \hline
                $D[\mathcal{A}']$ & 0.5        & 2.0        & 0.2           & 0.8           & $B\succ A$ \\
                $D$               & 1.31e-4    & 1.01e-7    & 0.9992        & 0.0008        & $A\succ B$ \\
                \hline
            \end{tabular}
        \end{center}
    \end{table}

    \emph{An important statement to prevent misinterpretation} is that although the above experiments may seem like a criticism of non-parametric hypothesis testing as well as Bayesian inference methods, this is \emph{not} the point of this paper.
    In fact, the statistical conclusions are based on a specific significance level and certain assumptions, and do not negate the possibility of opposing results.
    Thus, from a mathematical standpoint, these experiments—where special data were deliberately constructed to bypass the assumptions of the methodologies used—are somewhat meaningless.

    However, when we consider this issue more rigorously in the real world, we \emph{cannot} guarantee that real-world data will not exhibit behaviors similar to those in the examples.
    Without additional information and better processing methods, it is futile to apply statistical techniques to analyze available data based on certain assumptions.
    Yet, is it accurate to claim that “there is no more information and no better way to process the data”?
    Could we potentially “reduce errors due to mathematical assumptions by gathering more real data”?
    Can we provide more data for the specific scenario of algorithmic performance analysis to avoid or reduce the likelihood of the type II errors made by generic statistical methods?
    The next subsection will explore these questions further.

    \subsection{An analysis of the cause of the IIA paradox}\label{subsec:an-analysis-of-the-cause-of-the-iia-paradox}

    The previous experiments showed that certain statistical methods lead to the IIA paradox, as the comparison of Algorithms $A$ and $B$ is influenced by the inclusion or exclusion of Algorithms $C_2$ through $C_{98}$.
    This raises the question: how do Algorithms $C_2$ through $C_{98}$ impact the performance assessment of Algorithms $A$ and $B$?

    In non-parametric hypothesis testing, we compare Algorithms $A$ and $B$ based on their average rankings.
    In dataset $D[\mathcal{A}']$, the first 100 problems rank Algorithm $A$ one position higher than Algorithm $B$, while the last 400 problems rank Algorithm $A$ one position lower.
    However, numerically, Algorithm $A$ is significantly higher than Algorithm $B$ when ranked higher, whereas the difference is much smaller when Algorithm $A$ is ranked lower.
    This discrepancy is not captured when $D[\mathcal{A}']$ is considered in isolation.

    When Algorithms $C_2$ through $C_{98}$ are included in the set, the first 100 questions rank Algorithm $A$ 98 places higher than Algorithm $B$, while in the subsequent 400 questions, Algorithm $B$ ranks just one place higher than Algorithm $A$.\@
    This indicates that the inclusion of a third-party algorithm can significantly influence the rankings of $A$ and $B$.\@
    Notably, this effect occurs only when the performance of the third-party algorithm lies between that of Algorithms $A$ and $B$\@.

    \begin{figure}[!t]
        \centering
        \includegraphics[width=3.5in]{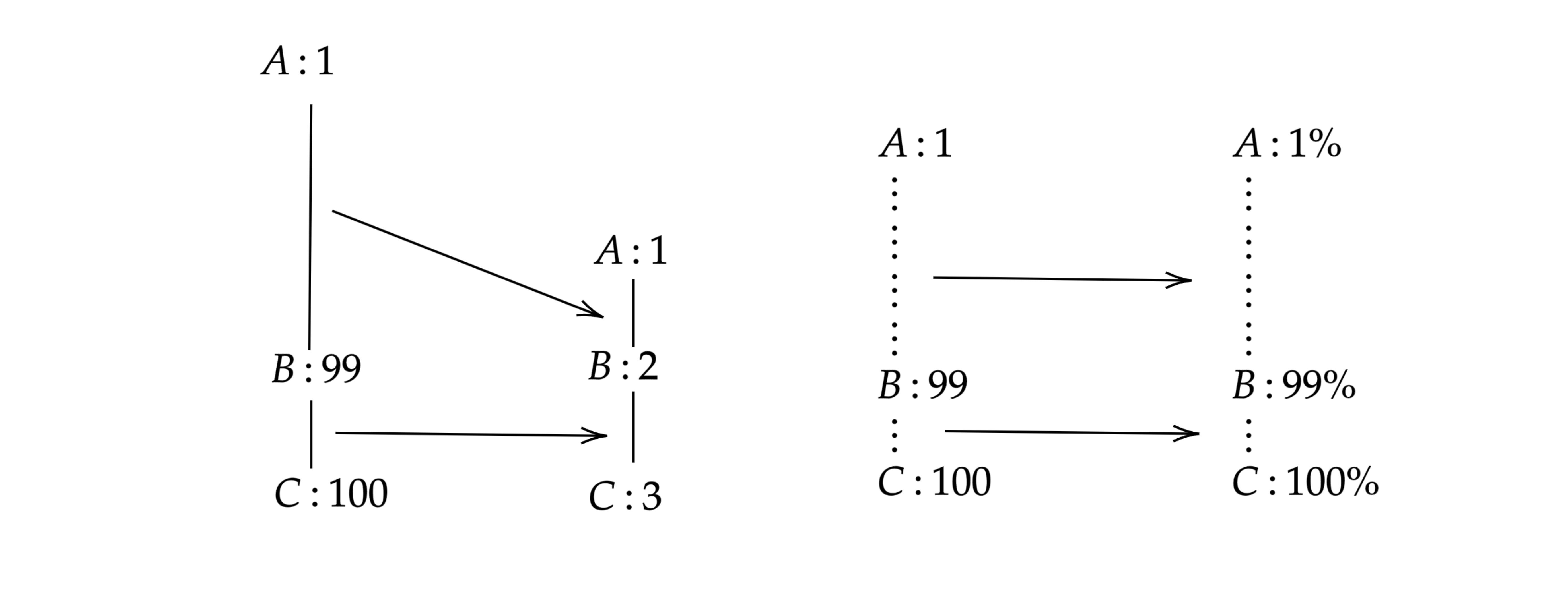}
        \caption{The erasure effect generated by using relative rankings.
        Algorithms positioned adjacently, regardless of the magnitude of their differences in raw data, have their gaps set to 1 after ranking.
        On the contrary, adding sufficient algorithms allows the performance differences between adjacent algorithms to become evident in their scores.}
        \label{fig:erase}
    \end{figure}

    Figure~\ref{fig:erase} visualizes this process.\footnote{This is just a schematic to show the effect of erasure, and the solution to the problem cannot be introduced by this schematic because it introduces another error, see subsection~\ref{subsec:a-response-idea-to-the-iia-paradox} for details.}
    When a small number of algorithms are involved in the comparison, the operation “find the rank of algorithm performance” \emph{squeezes} the performance difference between algorithms, no matter it is very large or very small, into a smaller performance difference.
    In this case, the relative ranking of the algorithms is not a good description of the performance quality of the algorithms.
    However, as the set of algorithms expands, the likelihood of third-party algorithms between algorithms with large performance differences is higher than the likelihood of them between algorithms with very small performance differences, which makes the average ranking of the algorithms closer to the original numerical performance.

    For better discussion, we refer to the process of ``replacing the actual performance results of algorithms with their performance ranking in the set of algorithms'' as ``relative ranking normalization'' and avoid lengthy descriptions by using the following definition:

    \begin{definition}[erasure effect]
        The erasure effect refers to a phenomenon in which data cannot be restored to its original value after processing.
    \end{definition}

    This paper does not imply that all erasure effects are harmful.
    For example, if the original data is disturbed by Gaussian noise, then the data processing of ``finding the mean'' produces an erasure effect that removes the Gaussian noise, which is beneficial.
    Therefore, the crux of the problem is whether the information erased by the data processing is the information that people really want to pay attention to.
    Obviously, relative ranking normalization with a small number of algorithms as the set of algorithms erases the actual amount of performance variance of the algorithms, and is therefore a bad erasure effect.

    The comparative conclusions of non-parametric hypothesis testing derive from average rankings, and the influence of third-party algorithms on these rankings can be intuitively understood.
    However, why do third-party algorithms affect Bayesian inference?
    We believe that by conducting an in-depth investigation into the computational process of Bayesian inference and carefully exploring the impact of incorporating third-party algorithms on its numerical outcomes, we can draw meaningful conclusions.
    Nonetheless, this paper presents a more sophisticated approach to the analysis.
    If we denote the data obtained by replacing each value in $D$ with its corresponding rank within its column as $R$, and the data representing the pairwise comparison results for Bayesian inference derived from the data $D$ as $B$, we will observe that $R$ and $B$ can be mutually transformed.
    Specifically, a sequence containing $n$ elements and the corresponding data with $n \choose 2$ pairwise comparison results can be converted between each other without loss, indicating that no erasure effect occurs in this convertion.
    This illustrates that during the processing of $D$ using Bayesian inference, a similar erasure effect to that observed in the transition from $D$ to $R$ occurs, and this information loss is irreversible.
    That answers why do third-party algorithms affect Bayesian inference.

    More generally, we propose an empirical condition for the occurrence of an IIA paradox:
    \begin{proposition}
        When an algorithmic comparison method produces the same results with $D$ as the data source as it does with $R$, there is a high probability that this algorithmic comparison method does not meet the IIA criterion.
    \end{proposition}

    When this condition is satisfied, this paper suggests further verification against the IIA paradox using $D$ and $D[\mathcal{A}']$ constructed in this paper.

    A discussion on how to reduce the erasure effect caused by relative ranking normalization is provided in the next subsection.

    \subsection{A response idea to the IIA paradox}\label{subsec:a-response-idea-to-the-iia-paradox}

    Before presenting the recommended solution in this paper, we first address a misconception that may arise from the misleading depiction in Figure~\ref{fig:erase}.
    Focusing on the right half of Figure~\ref{fig:erase}, one might suggest that max-min scaling resolves the issue illustrated.
    \begin{equation}
        \label{eq:mms}
        r = \frac{f_{max}-f}{f_{max}-f_{min}}
    \end{equation}
    where $f_{\max}$ and $f_{\min}$ are the maximum and minimum value of the objective function(or of the values in corresponding column in $D$ when they are unknown).

    However, in light of the isomorphism issue, max-min scaling only ensures compliance with the linear isomorphism criterion.
    It does not guarantee that Equation~\ref{eq:mms} will yield the same result across all functions within the isomorphism set.

    As demonstrated in the previous examples, one potential solution to mitigate the erasure effect caused by relative ranking normalization is to include as many third-party algorithms as possible, making the normalized data closer to the original.
    A theoretical approach is to add an infinite number of completely randomized algorithms—search algorithms that uniformly select points within the domain—resulting in a stable relative order that remains unaffected by a finite number of third-party algorithms.

    A deeper examination of this idea is beyond the scope of this paper and will be explored in a future paper, as fully addressing it could raise additional questions and divert attention from the main focus of this work.
    However, until the issue is more effectively addressed, we propose that expanding the algorithm set by incorporating as many algorithms as possible is a reasonable strategy.
    Additionally, reusing known algorithms with varied parameters and search strategies can significantly enhance the algorithm set.

    Although this method increases the workload of the experiment, it can theoretically reduce the probability of the IIA problem, and can also show the performance of the algorithms under different parameters and strategies more clearly, and also reduce the loss of the reliability of the conclusions due to the ``existence of the possibility of setting unreasonable parameters or strategies for other algorithms involved in the comparison'' to a certain extent.

    \section{Conclusions}\label{sec:conclusions}
    The main contribution of this paper is to propose two criteria for performance analysis of optimization algorithms.
    This includes: showing that the isomorphism criterion and the IIA criterion are proposed, verifying that there are commonly used performance evaluation metrics and algorithm comparison methods in practice that violate these two reasonable criteria, and providing methods for each criterion to assist in verifying whether the corresponding paradoxes occur.
    These two criteria provide validation standards for proposers of performance evaluation metrics and algorithm comparison methods.

    The secondary contribution of this paper is to analyze and propose solution ideas for the isomorphism paradox and the IIA paradox.

    At the methodological level, this paper emphasizes the combination of cross-disciplinary and practical issues, and advocates avoiding limiting the data sources for data analysis of algorithms to experimental results, as well as avoiding the simple application of other disciplines when choosing algorithmic ordering methods.

    In the future work, the solution ideas proposed in this paper will be further analyzed in depth and form corresponding solutions.

    \bibliographystyle{IEEEtran}
    \bibliography{main}

\end{document}